\documentclass[EPJ,twocolumn]{webofc}
\usepackage[varg]{txfonts}   % Web of Conferences font
\wocname{ISVHECRI 2012}
\woctitle{ISVHECRI 2012}
\begin{document}
\title{Measurement of Cosmic Ray Spectrum and Anisotropy with ARGO-YBJ}

\author{G. Di Sciascio\inst{1}\fnsep\thanks{\email{disciascio@roma2.infn.it} } on behalf of the ARGO-YBJ Collaboration }

\institute{INFN, Sezione di Roma Tor Vergata, Viale della Ricerca Scientifica 1, Roma, Italy I-00133.
          }

\abstract{%
The combined measurement of the cosmic ray (CR) energy spectrum and anisotropy in their arrival direction distribution needs the knowledge of the elemental composition of the radiation to discriminate between different origin and propagation models.
Important information on the CR mass composition can be obtained studying the EAS muon content through the measurement of the CR rate at different zenith angles.

In this paper we report on the observation of the anisotropy of galactic CRs at different angular scales with the ARGO-YBJ experiment.
We report also on the study of the primary CR rate for different zenith angles. 
The light component (p+He) has been selected and its energy spectrum measured in the energy range (5 - 200) TeV for quasi-vertical events. 
With this analysis for the first time a ground-based measurement of the CR spectrum overlaps data obtained with direct methods for more than one energy decade, thus providing a solid anchorage to the CR spectrum measurements carried out by EAS arrays in the knee region.

Finally, a preliminary study of the non-attenuated shower component at a zenith angle $\theta >$ 70$^{\circ}$ (through the observation of the so-called horizantal air showers) is presented.
} 
\maketitle
\section{Introduction}
\label{intro}

The understanding of CRs origin at any energy is made difficult by the poor knowledge of the elemental composition of the radiation. The determination of the CR arrival direction does not depend on knowledge of the mass of the primary particle, however, the use of combined data on the energy spectrum and arrival direction distribution requires the knowledge of the primary mass distribution to discriminate between different origin and propagation models.

Inclined showers ($\theta >$ 60$^{\circ}$) induced by very high-energy CRs are mainly produced by secondary muons, in contrast to the vertical ones dominated by photons and electrons stemming from $\pi^0$ decays. Measurements of the CR rate at different zenith angles gives information on the relative number of muons in a shower, which is dependent on the CR elemental composition, thus providing an important tool to probe the CR mass distribution.

As CRs are mostly charged nuclei, their arrival direction is deflected and highly isotropized by the action of galactic magnetic field (GMF) they propagate through before reaching the Earth atmosphere. 
However, different experiments \cite{nagashima,tibet06,eastop09,icecube11} observed an energy-dependent \emph{"large scale"} anisotropy (LSA) in the sidereal time frame with an amplitude of about 10$^{-4}$ - 10$^{-3}$, suggesting the existence of two distint broad regions, one showing an excess of CRs (called "tail-in"), distributed around 40$^{\circ}$ to 90$^{\circ}$ in Right Ascension (R.A.). The other a deficit (the so-called "loss cone"), distributed around 150$^{\circ}$ to 240$^{\circ}$ in R.A..

In the last years the Milagro \cite{milagro08} experiment reported evidence of the existence of a medium angular scale anisotropy (MSA) contained in the tail-in region.
The observation of similar small scale anisotropies has been recently claimed by the Icecube experiment in the Southern hemisphere \cite{icecube11}.

So far, no theory of CRs in the Galaxy exists which is able to explain the origin of these different anisotropies leaving the standard model of CRs and that of the local GMF unchanged at the same time.
A joint analysis of concurrent data recorded by different experiments in both hemispheres, as well as a measurement of energy spectrum and elemental composition of the anisotropy regions, should be a high priority to clarify the observations.

The ARGO-YBJ experiment, located at the YangBaJing Cosmic Ray Laboratory (Tibet, P.R. China, 4300 m a.s.l., 606 g/cm$^2$), is an air shower array able to detect the cosmic radiation with an energy threshold of a few hundred GeV. 
The full detector is in stable data taking since November 2007 with a duty cycle larger than 85\%. The trigger rate is 3.6 kHz. The detector characteristics are described in \cite{bacci00,aielli06,aielli09}.
Details on the analysis procedure (e.g., reconstruction algorithms, data selection, background evaluation, systematic errors) are discussed in \cite{aielli10,bartoli11}.
The performance (angular resolution, pointing accuracy, energy scale calibration) and the operation stability are monitored on a monthly basis by observing the Moon shadow, i.e., the deficit of CR detected in its direction \cite{bartoli11}. 
The last results obtained by ARGO-YBJ are summarized in \cite{gdisciascio12}.

In this paper the observation of CR anisotropy at different angular scales with ARGO-YBJ is reported for different primary energies.
We report also on the measurement of the primary CR spectrum for different zenith angles.

%%%%%%%%%%%%%%%%%%%%%%%%%%%%%%%%%%%%%%
\section{Cosmic Ray Anisotropy}
\label{sec-1}
%%%%%%%%%%%%%%%%%%%%%%%%%%%%%%%%%%%%%%

To study the anisotropy at different angular scales the isotropic background of CRs has been estimated with two well-known methods: the equi-zenith angle method \cite{amenomori05} and the direct integration method \cite{Fleysher}.
The equi-zenith angle method, used to study the LSA, is able to eliminate various spurious effects caused by instrumental and environmental variations, such as changes in pressure and temperature that are hard to control and tend to introduce systematic errors in the measurement.
The direct integration method, based on time-average, relies on the assumption that the local distribution of the incoming CRs is slowly varying and the time-averaged signal may be used as a good estimation of the background content. 
Time-averaging methods (TAMs) act effectively as a high-pass filter, not allowing to inspect features larger than the time over which the background is computed (i.e., 15$^{\circ}$/hour$\times \Delta t$ in R.A.) \cite{iupdis12}. The time interval used to compute the average spans $\Delta t$ = 3 hours and makes us confident the results are reliable for structures up to $\approx$35$^{\circ}$ wide. 

\subsection{Large Scale Anisotropy}
\begin{figure}
\centering
\sidecaption
\includegraphics[width=0.40\textwidth,clip]{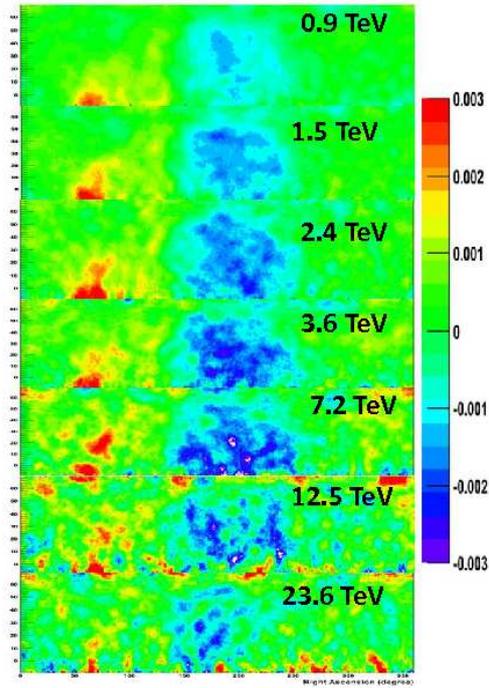}
\caption{Large scale CR anisotropy observed by ARGO-YBJ as a function of the energy. The color scale gives the relative CR intensity.}
\label{fig:lsa}       % Give a unique label
\end{figure}

The observation of the CR large scale anisotropy by ARGO-YBJ is shown in Fig. \ref{fig:lsa} as a function of the primary energy up to about 25 TeV. 

%
%%%%%%%%%%%%%%%%%%%%%%%%%%%%%%%%%%%%%%%%%%%%%%%%%%%%%%%%%%%%%%%%%
%%%%%%%%%%%%%%%%%%%%%%%%%%%%%%%%%%%%%%%%%%%%%%%%%%%%%%%%%%%%%%%%%
%
%%%%%%%%%%%%%%%%%%%%%%%%%%%%%%%%%%%%%%%%%%%%%%%%%%%%%%%%%%%%%%%%%
\begin{figure}
\centering
\sidecaption
\includegraphics[width=0.40\textwidth,clip]{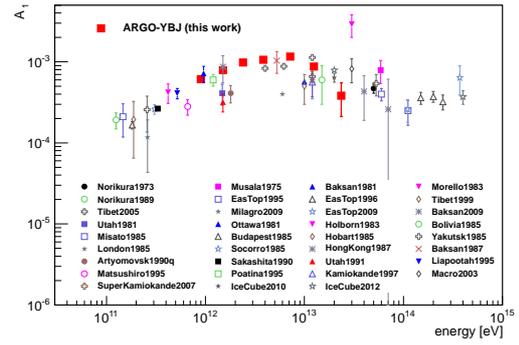}
\caption{Amplitude of the first harmonic as a function of the energy, compared with other measurements.}
\label{fig:lsa-ampl}       % Give a unique label
\end{figure}
%%%%%%%%%%%%%%%%%%%%%%%%%%%%%%%%%%%%%%%%%%%%%%%%%%%%%%%%%%%%%%%%%
%
%%%%%%%%%%%%%%%%%%%%%%%%%%%%%%%%%%%%%%%%%%%%%%%%%%%%%%%%%%%%%%%%%
\begin{figure}
\centering
\sidecaption
\includegraphics[width=0.40\textwidth,clip]{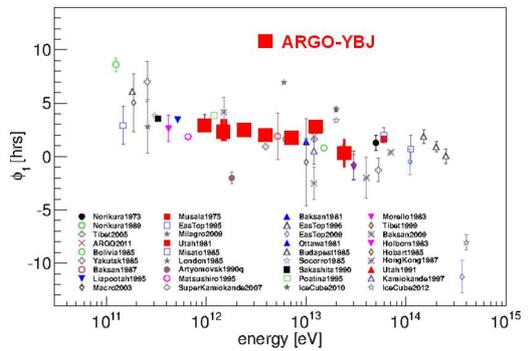}
\caption{Phase of the first harmonic as a function of the energy, compared with other measurements.}
\label{fig:lsa-phase}       % Give a unique label
\end{figure}
%%%%%%%%%%%%%%%%%%%%%%%%%%%%%%%%%%%%%%%%%%%%%%%%%%%%%%%%%%%%%%%%%
%
The so-called \textit{`tail-in'} and \textit{`loss-cone'} regions, correlated to an enhancement and a deficit of CRs, respectively, are clearly visible with a statistical significance greater than 20 s.d..
The tail-in broad structure appears to break up into smaller spots with increasing energy.
In order to quantify the scale of the anisotropy we studied the 1-D R.A. projections integrating the sky maps inside a declination band given by the field of view of the detector. Therefore, we fitted the R.A. profiles with the first two harmonics. The resulting amplitude and phase of the first harmonic are plotted in Fig. \ref{fig:lsa-ampl} and Fig. \ref{fig:lsa-phase} where are compared to a full compilation of measurements as a function of the energy. The ARGO-YBJ results are in agreement with those of other experiments, suggesting a decrease of the anisotropy first harmonic amplitude at energies above 10 TeV.
\subsection{Medium Scale Anisotropy}
\begin{figure}
\centering
\sidecaption
\includegraphics[width=0.40\textwidth,clip]{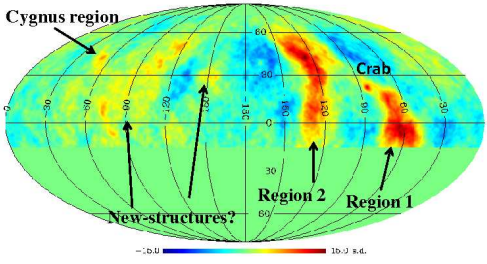}
\caption{Medium scale CR anisotropy observed by ARGO-YBJ. The color scale gives the statistical significance of the observation in standard deviations.}
\label{fig:msa}       % Give a unique label
\end{figure}

Fig. \ref{fig:msa} shows the ARGO-YBJ sky map in equatorial coordinates containing about 2$\cdot$10$^{11}$ events reconstructed with a zenith angle $\leq$50$^{\circ}$.
%
%%%%%%%%%%%%%%%%%%%%%%%%%%%%%%%%%%%%%%%%%%%%%%%%%%%%%%%%%%%%%%%%%
%%%%%%%%%%%%%%%%%%%%%%%%%%%%%%%%%%%%%%%%%%%%%%%%%%%%%%%%%%%%%%%%%
%
The zenith cut selects the declination region $\delta\sim$ -20$^{\circ}\div$ 80$^{\circ}$.
According to simulations, the median energy of the isotropic CR proton flux is E$_p^{50}\approx$1.8 TeV (mode energy $\approx$0.7 TeV).

The most evident features are observed by ARGO-YBJ around the positions $\alpha\sim$ 120$^{\circ}$, $\delta\sim$ 40$^{\circ}$ and $\alpha\sim$ 60$^{\circ}$, $\delta\sim$ -5$^{\circ}$, positionally coincident with the excesses detected by Milagro \cite{milagro08}. These regions, named ``1'' and ``2'', are observed with a statistical significance of about 14 s.d..  The deficit regions parallel to the excesses are due to a known effect of the analysis, that uses also the excess events to evaluate the background, overestimating this latter \cite{iupdis12}.

The left side of the sky map is full of few-degree excesses not compatible with random fluctuations (the statistical significance is more than 6 s.d.). 
The observation of these structures is reported here for the first time and together with that of regions 1 and 2 it may open the way to an interesting study of the TeV CR sky.
%
%%%%%%%%%%%%%%%%%%%%%%%%%%%%%%%%%%%%%%%%%%%%%%%%%%%%%%%%%%%%%%%%%
\begin{figure}
\centering
\sidecaption
\includegraphics[width=0.40\textwidth,clip]{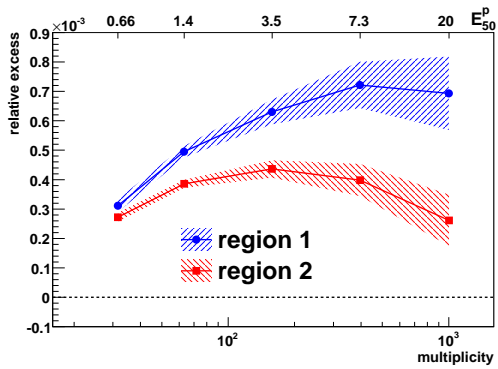}
\caption{Size spectrum of the regions 1 and 2. The vertical axis represents the relative excess (Ev-Bg)/Bg. The upper scale shows the corresponding proton median energy.}
\label{fig:msa-ensp}       % Give a unique label
\end{figure}
%%%%%%%%%%%%%%%%%%%%%%%%%%%%%%%%%%%%%%%%%%%%%%%%%%%%%%%%%%%%%%%%%
%

To figure out the energy spectrum of the excesses, data have been divided into five independent shower multiplicity sets. The number of events collected within each region are computed for the event map (Ev) as well as for the background one (Bg). The relative excess (Ev-Bg)/Bg is computed for each multiplicity interval. The result is shown in the Fig. \ref{fig:msa-ensp}. Region 1 seems to have spectrum harder than isotropic CRs and a cutoff around 600 shower particles (proton median energy E$^{50}_p$ = 8 TeV). On the other hand, the excess hosted in region 2 is less intense and seems to have a spectrum more similar to that of isotropic CRs.
As a reference value, the upper horizontal scale reports the median energy of isotropic CR protons for each multiplicity interval obtained via MC simulation. 

\subsection{The Compton-Getting effect}

\begin{figure}
\centering
\sidecaption
\includegraphics[width=0.37\textwidth,clip]{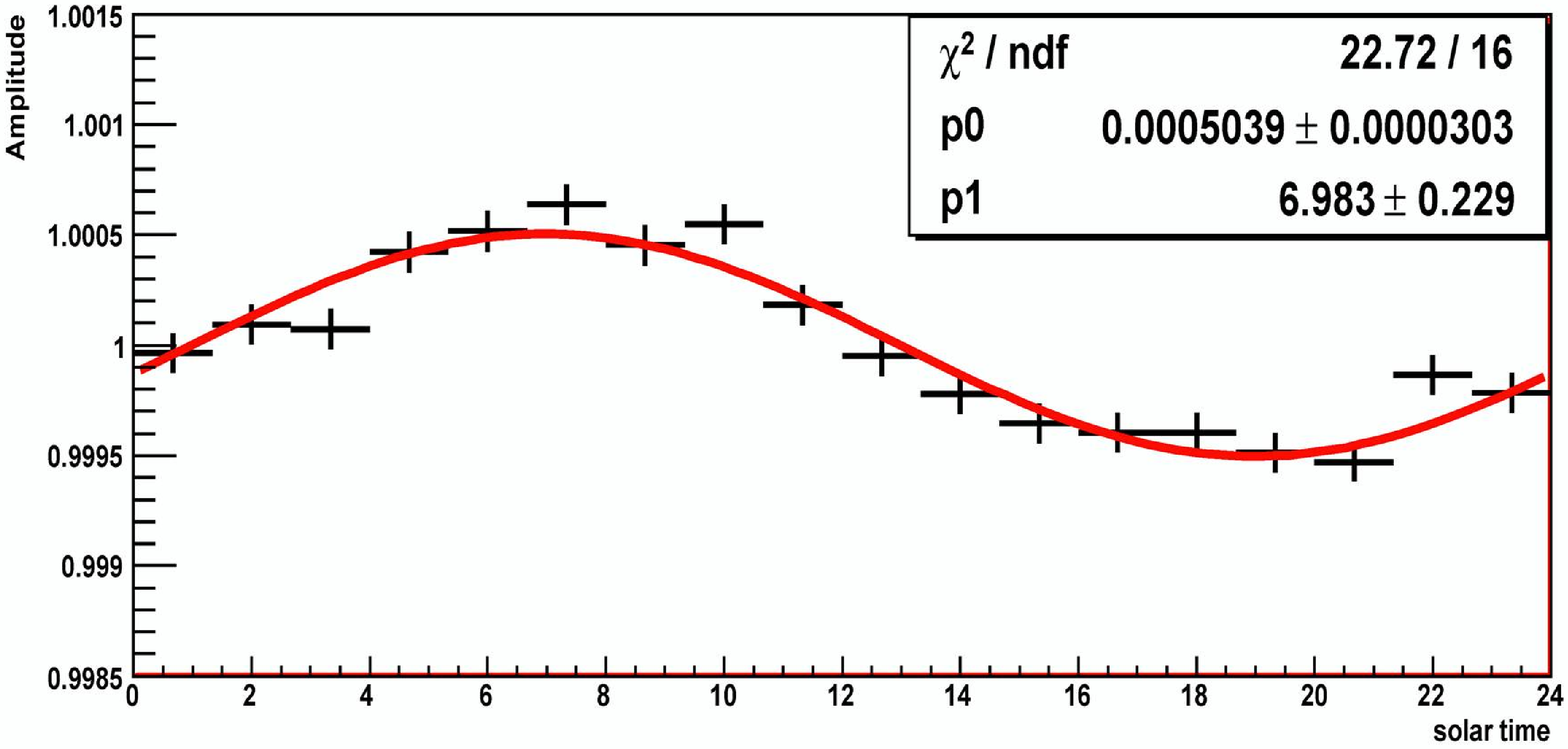}
\caption{One-dimensional projection in right ascension of the two-dimensional CR map in local solar time. The red line shows the best-fit to ARGO-YBJ data (crosses).}
\label{fig:comptgett}       % Give a unique label
\end{figure}

The origin of CR anisotropies is still unknown therefore, the observation of an expected anisotropy is important to check the reconstruction algorithms, the background calculation and the stability of the detector performance. A well-known expected anisotropy is the so-called Compton-Getting (CG) effect, a dipole anisotropy in the local solar frame, due to the Earth's motion around the Sun \cite{comptgett}. A significant signal compatible with CG is seen by ARGO-YBJ in solar time above $\sim$ 8 TeV to avoid additional effects due to heliospheric magnetic field and solar activity.
In fact, we found that including lower energy events results in much larger modulation amplitudes than those obtained
when these events were excluded.
Fig. \ref{fig:comptgett} shows the solar variations observed by ARGO-YBJ together with the sinusoidal curve best fitted to the data.
The fair agreement between data and calculations ($\phi$ = 6:00 hr, A = 9.7 $\cdot$ 10$^{-5}$) make us confident about the capability of ARGO-YBJ in detecting anisotropies at a level of 10$^{-4}$.  
%
%%%%%%%%%%%%%%%%%%%%%%%%%%%%%%%%%%%%%%%%%%%%%%%%%%%%%%%%%%%%%%%%%
%%%%%%%%%%%%%%%%%%%%%%%%%%%%%%%%%%%%%%%%%%%%%%%%%%%%%%%%%%%%%%%%
%

\section{CR primary spectrum}

\subsection{Light component (p+He) spectrum of CRs}
%
%%%%%%%%%%%%%%%%%%%%%%%%%%%%%%%%%%%%%%%%%%%%%%%%%%%%%%%%%%%%%%%%%
\begin{figure}
\centering
\sidecaption
\includegraphics[width=0.40\textwidth,clip]{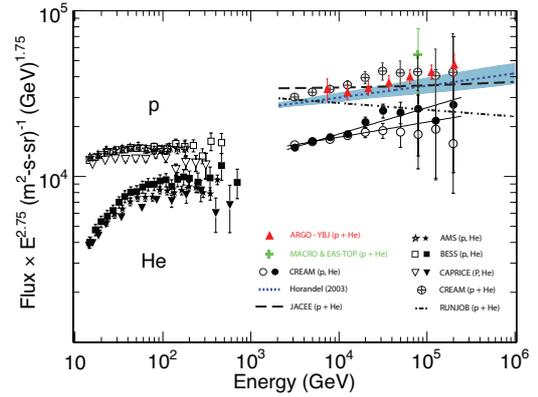}
\caption{Light component (p+He) energy spectrum of primary CRs measured by ARGO-YBJ compared with other experimental results.}
\label{fig:light_spectrum}       % Give a unique label
\end{figure}
%%%%%%%%%%%%%%%%%%%%%%%%%%%%%%%%%%%%%%%%%%%%%%%%%%%%%%%%%%%%%%%%
%
Requiring quasi-vertical showers ($\theta$ $<$ 30$^{\circ}$) and applying a selection criterion based on the particle density, a sample of events mainly induced by protons and helium nuclei, with shower core inside a fiducial area (with radius $\sim$28 m), has been selected. The contamination by heavier nuclei is found negligible. An unfolding technique based on the Bayesian approach has been applied to the strip multiplicity distribution in order to obtain the differential energy spectrum of the light component (p + He nuclei) in the energy range (5 - 200) TeV \cite{bartoli12}. 
The spectrum measured by ARGO-YBJ is compared with other experimental results in Fig. \ref{fig:light_spectrum}.  
Systematic effects due to different hadronic models (Corsika 6.710 with QGSJet-II and SYBILL) and to the selection criteria do not exceed 10\%.
The ARGO-YBJ data agree remarkably well with the values obtained by adding up the p and He fluxes measured by CREAM both concerning the total intensities and the spectral index \cite{cream11}. The value of the spectral index of the power-law fit to the ARGO-YBJ data is -2.61$\pm$0.04, which should be compared with $\gamma_p$ = -2.66$\pm$0.02 and $\gamma_{He}$ = -2.58$\pm$0.02 obtained by CREAM.
The present analysis does not allow the determination of the individual p and He contribution to the measured flux, but the ARGO-YBJ data clearly exclude the RUNJOB results \cite{runjob}. 
We emphasize that for the first time direct and ground-based measurements overlap for a wide energy range thus making possible the cross-calibration of the different experimental techniques.

\subsection{Horizontal Air Showers}
%
%%%%%%%%%%%%%%%%%%%%%%%%%%%%%%%%%%%%%%%%%%%%%%%%%%%%%%%%%%%%%%%%%
\begin{figure}
\centering
\sidecaption
\includegraphics[width=0.40\textwidth,clip]{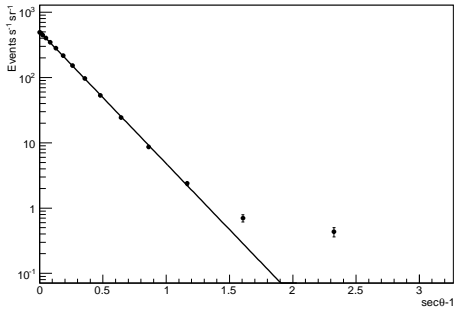}
\caption{The zenith angle distribution of EAS measured with ARGO-YBJ. The best fit out to $\sim$60$^{\circ}$ is also shown. }
\label{fig:has-secth}       % Give a unique label
\end{figure}
%%%%%%%%%%%%%%%%%%%%%%%%%%%%%%%%%%%%%%%%%%%%%%%%%%%%%%%%%%%%%%%%%
%
%%%%%%%%%%%%%%%%%%%%%%%%%%%%%%%%%%%%%%%%%%%%%%%%%%%%%%%%%%%%%%%%
\begin{figure}
\centering
\sidecaption
\includegraphics[width=0.40\textwidth,clip]{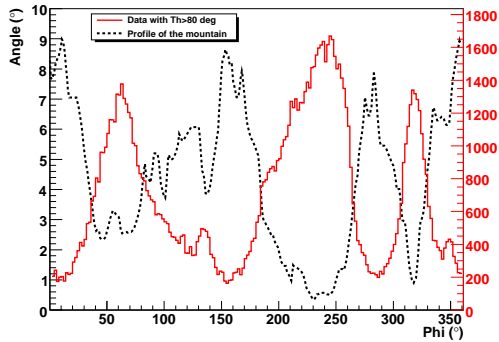}
\caption{Azimuthal distribution of showers with a reconstructed zenith angle $> 80^{\circ}$ (red dashed line) compared to the mountain profile seen by ARGO-YBJ (black continuous line).}
\label{fig:has-azimuth}       % Give a unique label
\end{figure}
%%%%%%%%%%%%%%%%%%%%%%%%%%%%%%%%%%%%%%%%%%%%%%%%%%%%%%%%%%%%%%%%
%
%%%%%%%%%%%%%%%%%%%%%%%%%%%%%%%%%%%%%%%%%%%%%%%%%%%%%%%%%%%%%%%%%
\begin{figure}
\centering
\sidecaption
\includegraphics[width=0.40\textwidth,clip]{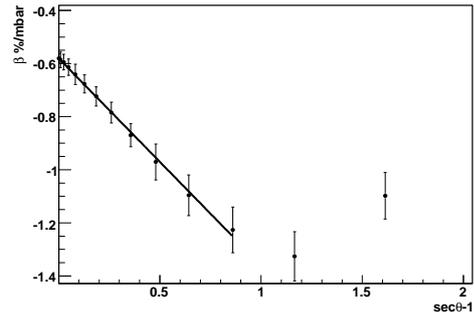}
\caption{The barometric coefficient for different zenith angles as measured by ARGO-YBJ.}
\label{fig:has-barom}       % Give a unique label
\end{figure}
%%%%%%%%%%%%%%%%%%%%%%%%%%%%%%%%%%%%%%%%%%%%%%%%%%%%%%%%%%%%%%%%
%
%
%%%%%%%%%%%%%%%%%%%%%%%%%%%%%%%%%%%%%%%%%%%%%%%%%%%%%%%%%%%%%%%%
\begin{figure}
\centering
\sidecaption
\includegraphics[width=0.40\textwidth,clip]{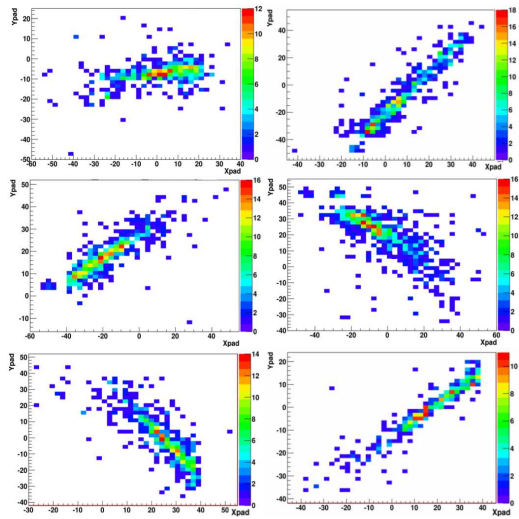}
\caption{Events observed by ARGO-YBJ with a reconstructed zenith angle $\theta >70^{\circ}$. Only showers with more than 500 fired strips on the central carpet are shown. The pixels represent 4$\times$4 pads (about 2$\times$2 m$^2$).}
\label{fig:has-ev}       % Give a unique label
\end{figure}
%%%%%%%%%%%%%%%%%%%%%%%%%%%%%%%%%%%%%%%%%%%%%%%%%%%%%%%%%%%%%%%%%
%

At zenith angles $\theta >$ 60$^{\circ}$ an excess of events (the so-called HAS, horizontal air showers) is observed above the rate of EAS as expected from the exponential absoprtion (with $\Lambda_{EAS} \approx$ 220 g/cm$^2$) of the air shower electromagnetic component in the large atmospheric depth (see Fig. \ref{fig:has-secth}), which implies a decrease of the EAS counting rate with $\Lambda_c \approx$ 130 g/cm$^2$.

The physical nature of these showers is confirmed by the absence of events from the direction of the sky shaded by the mountains around the ARGO-YBJ detector, as can be seen in the Fig. \ref{fig:has-azimuth} where the shower rate as a function of the reconstructed azimuthal angle is compared to the shadow angle due to the surrounding mountains. The expected anti-correlation is clearly visible and the mountain profile is reproduced quite well. 

Moreover, the dependence of the barometric effect on the zenith angle, shown in Fig. \ref{fig:has-barom}, clearly shows a deviation from the sec$\theta$ behaviour for sec$\theta >2$. In fact, the barometric coefficient $\beta = \frac{1}{n}\frac{dn}{dx}$ ($n$ = counting rate, $x$ = atmospheric pressure) is related to zenith angle as: $\beta(\theta) = \beta(0^{\circ}) sec\theta$. This can be explained by the presence of a "non-attenuated" EAS component that dominates for angles larger than 70$^\circ$.
Due to the small ARGO-YBJ effective area at large zenith angles, we expect that the observed HAS are due to high energy single muons which interact through bremmstrahlung (which dominate 10:1) or deep inelastic scattering and initiate showers at the appropriate depth (few hundreds g/cm$^2$) for detection, as shown in Fig. \ref{fig:has-ev} where some typical events observed by ARGO-YBJ are displayed. The characteristic elliptical shape of the showers, well contained in the central carpet, is clearly visible.
Such showers are essentially electromagnetic, since the remnant muons from the initial showers are dispersed over a very large area.
%
%%%%%%%%%%%%%%%%%%%%%%%%%%%%%%%%%%%%%%%%%%%%%%%%%%%%%%%%%%%%%%%%
\begin{figure}
\centering
\sidecaption
\includegraphics[width=0.40\textwidth,clip]{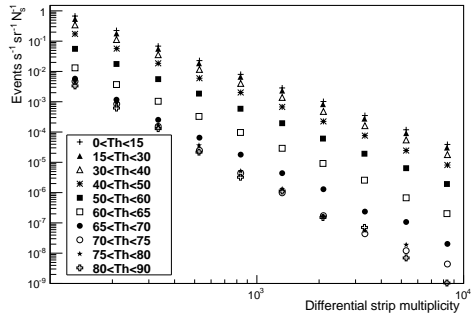}
\caption{Differential strip spectra measured by ARGO-YBJ for different zenith angles.}
\label{fig:has-spect}       % Give a unique label
\end{figure}
%%%%%%%%%%%%%%%%%%%%%%%%%%%%%%%%%%%%%%%%%%%%%%%%%%%%%%%%%%%%%%%%%%
%%%%%%%%%%%%%%%%%%%%%%%%%%%%%%%%%%%%%%%%%%%%%%%%%%%%%%%%%%%%%%%%%
\begin{figure}[ht!]
\centering
\sidecaption
\includegraphics[width=0.40\textwidth,clip]{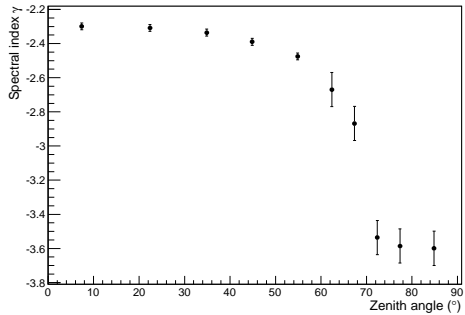}
\caption{Best-fit spectral indices calculated for the spectra of Fig. \ref{fig:has-spect}.}
\label{fig:has-indspect}       % Give a unique label
\end{figure}
%%%%%%%%%%%%%%%%%%%%%%%%%%%%%%%%%%%%%%%%%%%%%%%%%%%%%%%%%%%%%%%%%%

In Fig. \ref{fig:has-spect} the shower rate measured by ARGO-YBJ is shown, as a function of the fired strips number, for different primary zenith angles. The spectra soften with increasing angle up to about 70$^{\circ}$, as can be appreciated in Fig. \ref{fig:has-indspect} where the best-fit spectral indices are plotted. In the zenith angle region 50$^{\circ}$ - 70$^{\circ}$ a quick transition to a value of about -3.6, characteristic of the EAS muon component, is observed. 
Detailed simulations to reproduce the observed size spectrum of HAS are under way.

\section{Conclusions}

The ARGO-YBJ detector exploiting the full coverage approach and the high segmentation of the readout is imaging the front of atmospheric showers with unprecedented resolution and detail. The digital and analog readout will allow a deep study of the CR phenomenology in the wide TeV - PeV energy range. 

In this paper the observation of CR anisotropy at different angular scales is reported for different primary energies.
The large scale CR anisotropy has been clearly observed up to about 25 TeV.
The existence of different few-degree excesses in the Northern sky is showed by ARGO-YBJ for the first time.

We reported also on the measurement of the primary CR rate for different zenith angles. 
The light component (p+He) has been selected and its energy spectrum measured in the energy range (5 - 200) TeV for quasi-vertical events.
 
A preliminary study of HAS with ARGO-YBJ is also presented. More than 10$^7$ well-contained horizontal events have been analyzed, thus providing a ``well shielded'' sample useful to study the production and interaction of high energy CR muons and neutrinos.

%
% BibTeX or Biber users please use (the style is already called in the class, ensure that the "woc.bst" style is in your local directory)
% \bibliography{name or your bibliography database}

\end{document}